\documentclass[12pt,a4paper]{article}
\usepackage{amssymb}
\usepackage[T2A]{fontenc}
\usepackage[cp866]{inputenc}
\usepackage[russian,english]{babel}
\usepackage{graphicx}
\usepackage[dvips]{psfrag}
\newcommand{\risheight}{10cm}
\usepackage[unicode]{hyperref}
\begin{document}
\begin{center}
{\Large Correspondence between the physics of extremal black holes \\[3mm]
and that of stable heavy atomic nuclei}\\[3mm]
{B.~P.~Kosyakov${}^{a,b}$, E.~Yu.~Popov${}^a$, and  M. A. Vronski{\u\i}${}^{a,c}$}\\[3mm]
{{\small ${}^a$Russian Federal Nuclear Center--VNIIEF, 
Sarov, 607188 Nizhni{\u\i} Novgorod Region, Russia;\\
${}^b$Moscow Institute of Physics {\&} Technology, Dolgoprudni{\u\i}, 141700 Moscow Region, 
Russia;\\ 
${}^c$Sarov Institute of Physics {\&} Technology, Sarov, 607190 Nizhni{\u\i} Novgorod Region, Russia.}\\
{\tt E-mail:} 
${\rm  kosyakov.boris@gmail.com}$  (corresponding author)
} 
\end{center}
\begin{abstract}
\noindent
{
Extremal black holes are immune of Hawking evaporation.
On the other hand, some heavy atomic nuclei feature extraordinary stability to 
spontaneous transmutations changing their mass numbers.
The fact that extremal black holes and stable nuclei share a common trait, that 
of defying spontaneous ejection of their constituents, suggests that a good part 
of nuclear physics is modelled on physics of extremal black holes through a 
simple version of gauge/gravity duality. 
A general criterion for discriminating between stable and unstable microscopic 
systems can be formulated to gain a new insight into some imperfectly understood 
phenomena, such as instability of truly neutral spinless particles (Higgs bosons, 
$\pi_0$, quarkonia, glueballs).
}
\end{abstract}


\section{Introduction}
\label
{Introduction}
The conjectured equivalence between a quantum theory of gravity in 
anti-de Sitter space and quantum field theories in Minkowski space, known as 
``AdS/CFT'', or {``holography''}, or {``gauge/gravity duality''}
\cite{Maldacena}, \cite{Witten}, \cite{Gubser}, is generally believed to be a 
promising approach to both quantizing gravity and understanding the confinement of
quarks and gluons in the low energy limit of quantum chromodynamics (QCD).
A plausible assumption, advocated in this paper, is that there exist a 
holographic correspondence between extremal black holes and stable heavy atomic 
nuclei. 
To see this we use the following line of reasoning.

Black holes evaporate by Hawking radiation which owes its origin to the
feasibility of creations and annihilations of pairs of particles and 
antiparticles near the event horizon of the black holes. 
The sole exception are extremal black holes, they do not evaporate.
When isolated from other matter, they remain eternal.
The absence of Hawking radiation implies that relativistic quantum effects are 
suppressed.
The regime of evolution of extremal black holes is not only 
{\it semiclassical}---which yet allows creations and annihilations of particles
near the event horizon---but also {\it feeble quantum}, that is, immune of such processes.  

The situation closely parallels that in nuclear physics.
Certain of heavy nuclei feature extraordinary stability to spontaneous 
transmutations changing their mass numbers.
This property of heavy nuclei became pressing with the advent of QCD.
Indeed, it is no wonder that each of $3{\cal A}$ quarks, assembled into a nucleus, 
is individually kept in this nucleus from escaping.
However, {\it colorless clusters} of quarks and gluons, such as nucleons, light 
nuclei, 
and glueballs, are also permanently trapped in a stable heavy nucleus and unable 
for spontaneous detaching from it.
Note also that some properties of nuclei are typical of {\it classical} objects.
The most eloquent example is provided by the experimentally well established 
relationship between the size of a nucleus $R$ and its mass number ${\cal A}$:  
\begin{equation}
{R}={R_0}\, {\cal A}^{1/3}\,.
\label
{R-nucleus}
\end{equation}                                           
This relationship is inherent in a classical liquid drop rather than a 
quantum-mechanical system whose extension given by its Compton wavelength is 
inversely proportional to ${\cal A}$.

The fact that extremal black holes and stable heavy nuclei share a common trait, that of defying 
spontaneous ejection of their constituents 
\footnote{To make matters as simple as possible, we ignore the electromagnetic 
and weak couplings of quarks, so that the effects of $\gamma$ and $\beta$ 
emissions by nuclei are beyond the scope of the present discussion.
In this connection, the parallels between black hole evaporations and spontaneous ejections of heavy 
constituents of nuclei, such as glueballs and fission fragments, may seem 
far-fetched.
However, what counts is that, in the closing stages of evaporation, the black 
hole spectrum contains both light and {heavy} particles.}, suggests that these 
systems, governed by {semiclassical, feeble quantum}, laws of evolution, are 
related by a peculiar form of gauge/gravity duality.
What is the special feature of this duality?

Let us take a look at the holographic mapping in the general case.
A convenient coordinate patch, the Poincar\'e patch, covering one-half the $d$-dimensional anti-de 
Sitter space (AdS${}_d$ for short) gives the coordinatization with the metric
\begin{equation}
ds^2=\frac{L^2}{z^2}\left(d\tau^2-\sum_{i=1}^{d-2}dx_i^2-dz^2\right), 
\label
{Poincare_wedge}
\end{equation}                                           
where $z\in[0,\infty )$.
Upon an Euclidean continuation of AdS${}_d$, the boundary of AdS${}_d$ is 
${\mathbb E}_{d-1}$ at $z=0$ and a single point at $z=\infty$.  
The basic prescription for the evaluation of the desired mapping  \cite{Witten}, 
\cite{Gubser} is to identify the generating functional for $(d-1)$-dimensional 
Euclideanized Green's functions in the gauge theory $W_{\rm gauge}$ with its 
$d$-dimensional dual $Z_{\rm gravity}$ subject to the boundary conditions that 
a field $\Psi$ involved in both holographic sides becomes $\Psi(x,z=0)=\Psi(x)$
at $z=0$,
\begin{equation}
Z_{\rm gravity}[\Psi(x)]=W_{\rm gauge}[\Psi(x)]\,. 
\label
{basic_prescription}
\end{equation}                                           
We take $\Psi$ to be a {\it Dirac field}.
In the gauge side,  $\Psi$ is associated with the {\it quark field} appearing in an 
effective theory to low energy QCD.

We restrict our attention to  AdS${}_5$.
We use the metric signature $(+1,-1,-1,-1,-1)$, because this sign convention is best suited for the 
treatment of spinors, and we take notations and conventions related to the 5-dimensional Dirac 
equation which were adopted in Ref.~\cite{Wu}.
The 5-dimensional Dirac action in a black hole background \cite{Wu} reads 
\begin{equation}
S=\int d^5x\,\sqrt{-g}\,{\Psi}^\dagger\left[
\gamma^A e^\mu_A\left(\partial_\mu+\Gamma_\mu-ieA_\mu\right)+im
\right]\Psi\,.
\label
{EMD_action}
\end{equation}                                           
Here, $\Psi$ is a four-component Dirac spinor, $e^\mu_A$ is a pentad, $\Gamma_\mu$ is the spinor 
connection.
$A_\mu$ denotes the 5-dimensional vector potential. 
We choose units in which $\hbar$, $c$, and $G_{(5)}$ are unity. 
The set of matrices $\gamma^A$ is spanned by the quartet of Dirac $4\times 4$-matrices and $\gamma^5$, 
which realize the 5-dimensional Clifford algebra, $\{\gamma^A,\gamma^B\}=2\eta^{AB}$. 
Latin letters $A,B$ denote local orthonormal Lorentz frame indices $0,1,2,3,5$, while Greek letters 
$\mu,\nu$ run over five indices of spacetime  coordinates. 
The 5-dimensional Clifford algebra has two reducible representations, so that the Dirac field in 
(\ref{EMD_action}) can be treated in the 4-dimensional context, with $\gamma^5$ being the fifth 
basis vector component, and the spinor connection is given by a so$(1,4)$-valued 1-form 
$\Gamma_A=e_A^\mu\Gamma_\mu=\frac14\gamma^B\gamma^Cf_{BCA}$, where $f_{BCA}$ stands for the
structure constants of so$(1,4)$. 

Our main concern here is with the holographic image of extremal Reissner--Nordstr{\o}m 
black holes in AdS${}_5$ \footnote{In fact, we have no prior knowledge of the 
gravity theory dual to the physics of stable nuclei.
We therefore begin with the simplest descendant of the type II superstring theory.}.
To adapt the basic prescription (\ref{basic_prescription}) to semiclassical, 
feeble quantum, dynamics governing the behavior of extremal black holes we put 
\begin{equation}
Z_{\rm gravity}\sim e^{-{\bar S}[\Psi(x)]}\,, 
\label
{semiclassic_prescription}
\end{equation}                                           
where ${\bar S}[\Psi(x)]$ is an Euclideanized extremum of the action (\ref{EMD_action}) as a 
functional of $\Psi(x)$.
This is the same as saying the wave function $\Psi(x)$ of a Dirac particle in 
the AdS${}_5$ bulk is described by a solution to the Dirac equation
\begin{equation}
\left[\gamma^A e^\mu_A\left(\partial_\mu+\Gamma_\mu-ieA_\mu\right)+im\right]\Psi=0\,, 
\label
{Dirac_bulk}
\end{equation}                                           
where $\Gamma_A$ and $A_A$ represent gravitational and electromagnetic 
backgrounds of black holes.

The 4-dimensional semiclassical, feeble quantum, picture offers what amounts to 
its 5-dimensional dual.
However, if we are to think of the former as an effective theory in the infrared, 
all irrelevant degrees of freedom must be integrated out, except for degrees of 
freedom of a single quark $Q$ specified by the Dirac field $\Psi$.
This quark is affected by the mean field generated by all other constituents of 
a given many-quark system.
The 4-dimensional dynamics of this quark is assumed to be defined by the action
\begin{equation}
{\cal S}
=\int d^4x\left\{{\Psi}^\dagger\left[\gamma^\alpha\left(i\partial_\alpha+g_VA_\alpha\right)-m\right]\Psi 
+g_S{\Psi}^\dagger\Psi\,\Phi\right\}.
\label
{QCD-Lagrangian}
\end{equation}                                           
Here, $A_\alpha=(A_0,-{\bf A})$ and $\Phi$ are respectively the Lorentz vector potential and Lorentz 
scalar potential 
\footnote{The field $\Phi(x)$ is absent from the fundamental QCD Lagrangian because the 
scalar Yukawa coupling is contrary to asymptotic freedom, but the effective 
dynamics  in low energy region is anticipated to arrange itself into the form 
shown in Eq.~(\ref{QCD-Lagrangian}).} 
of the mean field, $g_V$ and $g_S$ are their associated couplings, and $m$ is the current-quark mass 
of the quark $Q$. 

Just as an {\it extremal} path contribution {\it dominates} the semiclassical, 
feeble quantum, {path integral} for the partition function in the bulk, Eq.~(\ref{semiclassic_prescription}), 
so does its dual on the screen, 
\begin{equation}
W_{\rm gauge}\sim e^{-{\bar{\cal S}}[\Psi(x)]}\,. 
\label
{semiclassic_prescription_s}
\end{equation}                                           
Here, ${\bar{\cal S}}[\Psi(x)]$ is an extremal value of the Euclideanized action (\ref{QCD-Lagrangian}) 
regarded as a functional of $\Psi(x)$,  the wave function of a single quark $Q$ incorporated 
into some nucleus.
Therefore, $\Psi(x)$ is given by solutions to the Dirac equation in the classical background 
$A_\alpha(x)$ and $\Phi(x)$ representing the mean field generated by all constituents of the 
nucleus. 

We thus see that the essentials of the present gauge/gravity correspondence are 
greatly {simplified} as against those in the general case.
We need no go into calculations of the connected Euclidean Green's functions of a 
gauge theory operator appearing in the basic prescription formulated in 
\cite{Witten}, \cite{Gubser}, because 
$\Psi$ is the one-particle {\it wave function}, rather than a quantized field.  
The level  of description is {relegated from quantum field theory} where 
creations and annihilations of quark-antiquark pairs are of major importance to 
{\it nonrelativistic quantum mechanics} in which the probability of these 
processes is negligible.
It will suffice to relate distinctive characteristics of a solution to the Dirac equation in 
the gravitational and electromagnetic background of an extremal black hole to those of the 
pertinent solution to the Dirac equation for a single quark $Q$ moving in the mean field of a
stable nucleus.

The idea that a single quark $Q$ driven by the mean field of its nucleus (or, alternatively, free 
nucleon) is responsible for static properties of this nucleus (free nucleon) turns out to be 
appropriate \cite{KPV}, \cite{KPV-2}.
However, the key premises of the analysis proposed in \cite{KPV} and \cite{KPV-2}, 
the pseudospin symmetry condition (or, alternatively, spin symmetry condition)~\footnote{For an extended 
discussion of these symmetries see Refs.~\cite{Ginocchio} 
and \cite{Liang}.} and growing mean field potentials, are phenomenological in 
nature.

It transpires that the direct implication of these premises---a spherical cavity
to which the quark $Q$ is permanently confined---arises quite naturally, 
that is, from the fundamental laws of gravitation and electromagnetism, in 
the holographic approach to the description of stable heavy nuclei.

It may be worth noting that AdS/CFT correspondence \cite{Maldacena}, \cite{Witten}, 
\cite{Gubser} is a {\it conjecture}.
This correspondence in its original form bears no relation to physics in our four-dimensional 
world.
Therefore, to obtain a ``realistic'' gauge/gravity duality, it is necessary to 
involve additional assumptions, which implies that {\it any} holographic model might be blamed 
for being {\it speculative}.
Meanwhile, a profound way to justfy the idea of holography is to reveal a clear and tenable mapping 
between some quantum laws of our microscopic world and some laws of gravitation in the 
five-dimensional anti-de Sitter space.
The main purpose of the present paper is to show that there is a very simple mapping of this kind 
which holds much promise in nuclear physics applications.
   
The paper is organized as follows.
The treatment of nuclear physics in terms of quark degrees of freedom, developed 
in \cite{KPV} and \cite{KPV-2}, is briefly reviewed in Sec.~2.
We point out here that the stability of a heavy nucleus (free nucleon) 
has a 
direct bearing on the pseudospin symmetry condition (spin symmetry condition).
Section~3 outlines the properties of solutions to the 5-dimensional Dirac 
equation (\ref{Dirac_bulk}) in static extremal black hole geometries which 
provide insight into the gravitational analog of a combination of the 
pseudospin symmetry condition and the spin symmetry condition.
This subject is further refined in Sec.~4.
Section~5 summarizes the features of the present holographic correspondence.

\section{Nuclei in the low energy QCD context}
\label
{nuclei}
We begin with the Dirac equation resulting from the action (\ref{QCD-Lagrangian}). 
We restrict our attention to spherically symmetric static interactions, and assume that the 
contribution of the Lorentz vector potential to the mean field is given by $A_0$.
What this means is the quark $Q$ orbits the center of mass, being driven by central potentials 
$A_0(r)$ and $\Phi(r)$, and roams around the nucleus, that is, the quark $Q$ is affected not only by 
the neighbouring quarks of the ``parent'' nucleon, but the combined potentials of the entire nucleus.
The arguments in support of this assumption closely resemble those taken in the single-particle 
shell model of atomic nuclei \cite{KPV-2}.
We thus take, as the starting point, the Dirac Hamiltonian
\begin{equation}
H=-i{\bf\alpha}\cdot{\nabla}+U_V({r})+\beta [m+U_S({r})]\,,
\label
{Dirac-Hamiltonian}
\end{equation}
in which $U_V=g_VA_0$, $U_S=g_S\Phi$, and $m$ is the reduced mass. 

The form of $U_V$ and $U_S$ is conveniently fixed to be one-half the Cornell potential \cite{Cornell75} 
\begin{equation}
{V_{\rm C}(r)}=
-\frac{\alpha_s}{r}+\sigma r \,,
\label
{Cornell}
\end{equation}
which is particularly appealing for the quarkonium  phenomenology \cite{KPV-2}.

To proceed to  the eigenvalue problem for the Dirac Hamiltonian, 
\begin{equation}
H\Psi({\bf r})=\varepsilon\Psi({\bf r})\,,
\label
{Dirac-genera}
\end{equation}
we first separate variables in the usual fashion \cite{Ginocchio}.
The radial part of Eq.~(\ref{Dirac-genera}) is
\begin{equation}
{f'}+\frac{1+\kappa}{r}\,f-ag=0\,,
\label
{Dirac_radia_f}
\end{equation}
\begin{equation}
{g'}+\frac{1-\kappa}{r}\,g+bf=0\,,
\label
{Dirac_radia}
\end{equation}
where 
$\kappa=\pm(j+\frac12)$ are eigenstates of the operator ${K}=-\beta\,({\bf S}\cdot{\bf L}+1)$ which 
commutes with the spherically symmetric Dirac Hamiltonian \cite{Ginocchio}, and
\begin{equation}
a(r)=\varepsilon+{m}+U_S(r)-U_V(r)\,,
\label
{A-df}
\end{equation}
\begin{equation}
b(r)=\varepsilon-{m}-U_S(r)-U_V(r)\,.
\label
{B-df}
\end{equation}
We use (\ref{Dirac_radia_f}) for expressing $g$ in terms of $f$ and substitute the result in 
(\ref{Dirac_radia}).
We eliminate the first derivative of $f$ from the resulting second-order differential equation to 
obtain the Schr\"odinger-like equation
\begin{equation}
F''+k^2F=0\,,
\label
{1D_Schroedinger}
\end{equation}                                          
where
\begin{equation}
k^2={\varepsilon^2-m^2}-2U(r;\varepsilon)=-\frac12\,A'(r)-\frac14\,A^2(r)
+B(r)\,,
\label
{k-df}
\end{equation}
\begin{equation}
A=-\frac{a'}{a}+\frac{2}{r}\,,
\quad 
B=a\,({1+\kappa})\left(\frac{1}{{r}a}\right)'+ab+\frac{1-\kappa^2}{r^2}\,.
\label
{B_1-df}
\end{equation}
The component $f$ (rather than $g$) is the focus of attention, because it is $f$ that survives in 
the nonrelativistic free-particle limit.

We take the pseudospin symmetry condition 
\begin{equation}
U_S=-U_V+C_{c}\,,
\label
{PSSC}
\end{equation}
where $C_{c}$ is a constant.
With Eq.~(\ref{PSSC}), the Dirac Hamiltonian (\ref{Dirac-Hamiltonian}) becomes
\begin{equation}
H={\bf\alpha}\cdot{\bf p}+U_V({r})(1-\beta)+ \beta(m+C_{c})\,.
\label
{Dirac-Hamiltonian-spin}
\end{equation}
We thus see that $m$ is shifted, 
\begin{equation}
m\to m_c=m+C_{c}\,.
\label
{mass_shift}
\end{equation}
The shift signals that the current quark mass converts to the corresponding 
constituent quark mass.
In what follows $m_c$ is regarded as the constituent quark mass of the quark $Q$ responsible for 
the static properties of nuclei, and the label $c$ of $m_c$ is omitted.
 
Let us degress briefly to notice that the description of free hadron states 
requires the spin symmetry condition  \cite{Ginocchio}, \cite{Liang},
\begin{equation}
U_{S}=U_V+{\bar{C}}_c\,.
\label
{SSC}
\end{equation}
Reasonably accurate results for the spectrum of quarkonia can then be found with 
the use of the discussed Dirac equation machinery \cite{KPV}, \cite{KPV-2}.

\begin{figure}[htb]
\centerline{\includegraphics[height=\risheight,angle=270]{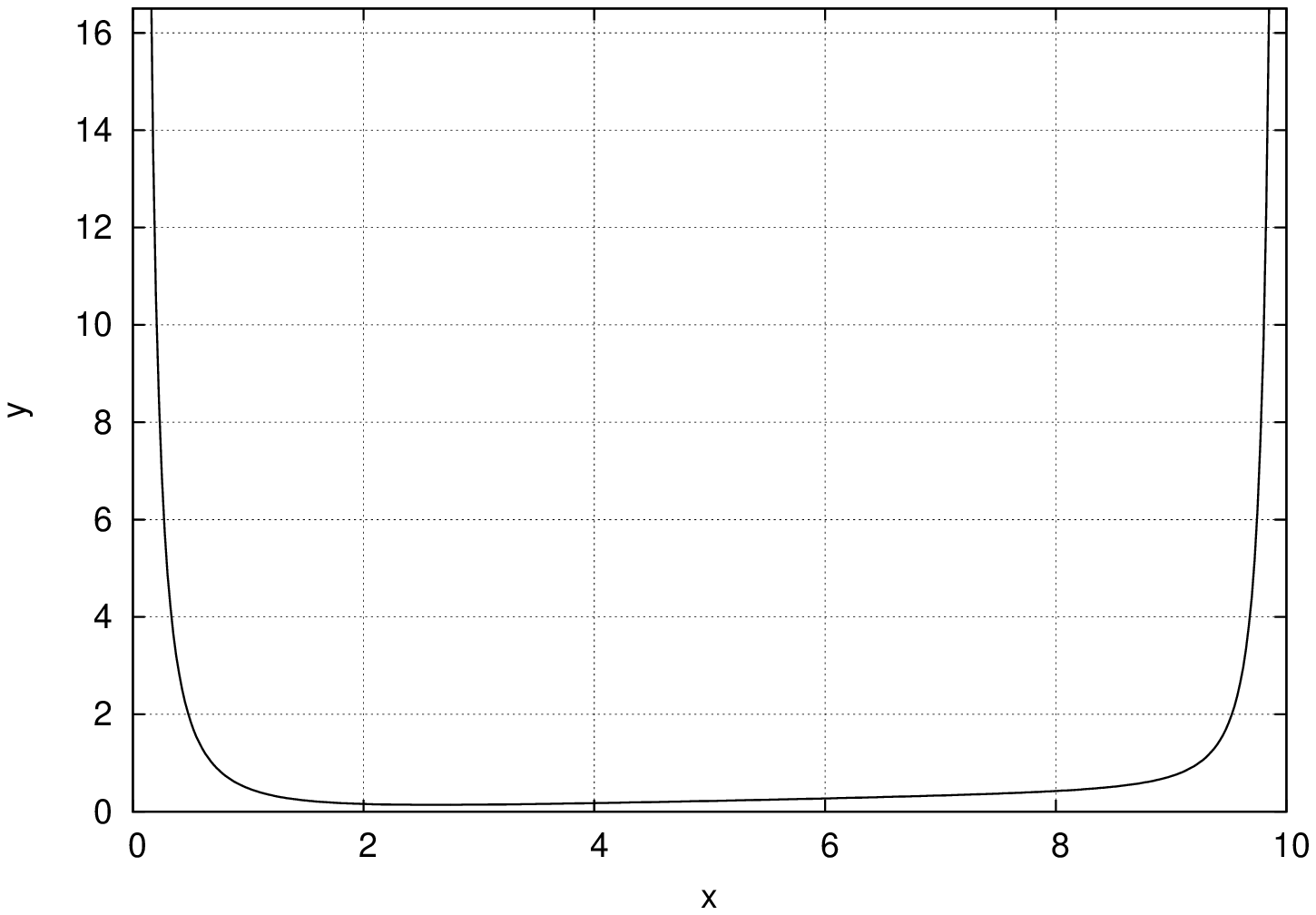}}
\caption{The effective potential (\ref{U_eff-PSEUDO}) with the parameters
$m=0.33\,{\rm GeV}$, $\varepsilon=1\,{\rm GeV}$, 
$\alpha_s=0.7$, $\sigma=0.14\,{\rm GeV}^2$, $\kappa=1$}\label
{pseudo-spin-potential}
\end{figure}

Taking $U_V=\frac12 V_{\rm C}$ and using (\ref{PSSC}) in (\ref{k-df}) and (\ref{B_1-df}), we find 
the effective potential
\[U(r;\varepsilon)=\frac{1}{2{r^2}}\Biggl\{{\kappa(\kappa+1)}
+\left(\varepsilon-{m}\right)\left(-\frac{\alpha_s}{r}
+\sigma r\right){r^2}
\]
\begin{equation}
+\frac{3(\alpha_s+\sigma r^2)^2}
{4\left[\sigma r^2-\left(\varepsilon +{m}\right)r-\alpha_s\right]^2}+
\frac{\alpha_s(\kappa+1)+\kappa\sigma r^2}
{\sigma r^2-\left(\varepsilon +{m}\right)r-\alpha_s}
\Biggr\}\,.
\label
{U_eff-PSEUDO}
\end{equation}
The last two terms of (\ref{U_eff-PSEUDO}) are singular at the point $r=r_{\rm sc}$ which is
the positive root of the equation $\sigma r^2-\left(\varepsilon +{m}\right)r-\alpha_s=0$,
\begin{equation}
r_{\rm sc}=\frac{\left(\varepsilon +{m}\right)+
\sqrt{\left(\varepsilon +{m}\right)^2+4\sigma \alpha_s}}{2\sigma}\,.
\label
{sol}
\end{equation}
The form of $U(r;\varepsilon)$ 
with particular values of $m$, $\varepsilon$, $\alpha_s$, $\sigma$, and 
$\kappa$ is depicted in Fig.~\ref{pseudo-spin-potential}. 
\psfrag{x}[c][c][0.7]{$r$,\quad GeV$^{-1}$}
\psfrag{y}[c][c][0.7]{$U(r;\varepsilon)$,\quad GeV}

The pseudospin symmetry condition (\ref{PSSC}) vastly enhances the 
interaction between the mean field and spin degrees of freedom of the quark $Q$ (more specifically,  
between the potential $V_{\rm C}$ and components $f$ and $g$ of the wave function $\Psi$) to yield 
a spherical shell of radius $r_{\rm sc}$ on which $U(r;\varepsilon)$ is infinite. 
The boundary of the spherical cavity of radius $r_{\rm sc}$ keeps the quark $Q$ in this cavity from 
escaping \cite{KPV}, \cite{KPV-2}.

A singular boundary arises whenever $U_V(r)$ grows indefinitely with $r$ because in going from 
Eqs.~(\ref{Dirac_radia_f}) and (\ref{Dirac_radia}) to Eq.~(\ref{1D_Schroedinger}) we have to apply 
the factor $1/a$ which is infinite when $a=0$.
The condition $U_S=-U_V$ implies that $a=\varepsilon+{m}-2U_V$, and $a=0$ has a positive root 
provided that $U_V$ increases monotonically with $r$ beginning at $r=0$ where $U_V$ assumes a 
negative value.
No singular boundary arises when $U_V\to U_0$ as $r\to\infty$, where $U_0$ is a 
constant which is less than $\frac12\left(\varepsilon+{m}\right)$. 
We are thus free to vary the form of the used potentials $U_V$ and $U_S$ in a wide range to attain the 
best fit to experiment.

It is reasonable to identify the spherical cavity with the interior of the nucleus over which the 
quark $Q$ executes periodic motions.
This identification gives a natural extension of the concept of confinement to nuclear physics: 
in the cavity, the probability amplitude of a quark contained in the nucleus is represented 
by solutions to the one-dimensional Schr\"odinger-like equation (\ref{1D_Schroedinger}), and
the probability amplitude to find this quark outside the cavity is zero \cite{KPV-2}.

To verify that the effective potential $U(r;\varepsilon)$ defined by Eq.~(\ref{U_eff-PSEUDO}) is 
indeed attributable to the description of stable heavy nuclei, that is, $3{\cal A}$-quark systems, 
we solved numerically Eqs.~(\ref{Dirac_radia_f}) and (\ref{Dirac_radia}) using the parameters 
$\alpha_s=0.7$ and $\sigma=0.1\,{\rm GeV}^2$ (borrowed from the description of quarkonia), and 
taking $m$ to be 0.33\,{GeV}.
The procedure was detailed in \cite{KPV-2}.
We found the energy levels $\varepsilon_{n_r}$ for $\kappa=-1,-2$ and the corresponding sizes of the 
cavities $r_{\rm sc}({n_r})$.
To a good approximation the energy levels $\varepsilon_{n_r}$ turn out to be proportional to 
$\sqrt{n_r}$ \cite{KPV-2}.
By assuming that  ${n_r}$ is proportional to ${\cal A}^{2/3}$, where ${\cal A}$ is the nucleus mass number, we came to
the relationship $r_{\rm sc}=R_0{\cal A}^{1/3}$ with $R_0\approx 1$ fm for the chosen values of the 
$\alpha_s$ and $\sigma$, which is consistent with Eq.~(\ref{R-nucleus}).
To verify that the assumption ${n_r}\sim{\cal A}^{2/3}$ is consistent with the experimental data, we compared 
the magnetic dipole of the quark $Q$ and that of the nucleus in which this quark is incorporated 
\cite{KPV-2}.
The agreement between our calculations and the observed values of the nuclear magnetic dipoles is 
for the most part within $\sim 20\%$.

\section{Dirac particles in charged static extremal black holes}
\label
{extreme_BH}
The 5-dimensional anti-de Sitter Reissner--Nordstr{\o}m geometry is described by
\begin{equation}
ds^2= h_t^2(r^2)\,dt^2-{h_t^{-2}(r^2)}{dr^2}-r^2d\Omega^2_{3}\,,
\label
{RN_AdS}
\end{equation}
where 
\begin{equation}
h_t^2(r^2)=1-\frac{2M}{r^2}+\frac{Q^2}{r^4}+\frac{r^2}{L^2}=
\frac{\Delta(r^2)}{L^2r^4}\,,
\label
{f(r)}
\end{equation}
\begin{equation}
\Delta(x)=x^3+L^2x^2-2L^2Mx+L^2Q^2\,,
\label
{Delta-r)}
\end{equation}
$d\Omega^2_{3}$ is the round metric in $S^3$,
\begin{equation}
d\Omega^2_{3}=d\psi^2+\sin^2\psi\left(d\vartheta^2+\sin^2\vartheta\,d\varphi^2\right),
\label
{dOmega}
\end{equation}
$M$ and $Q$ denote respectively the mass and electric charge of the 
hole,
$L$ is the curvature radius of AdS${}_5$.
The simplest solution to Maxwell's equations in this static manifold is 
$A_\mu=(A_0,0,0,0,0)$, where 
\begin{equation}
A_{0}(r)=\frac{Q}{r^2}\,.
\label
{A_and_A_5}
\end{equation}

Horizons of the metric are related to positive roots of $\Delta(x)$.
To find them, we define  
\begin{equation}
D=p^3+q^2\,,
\label
{D_df}
\end{equation}
where
\begin{equation}
p=-\frac{L^2}{3}\left(\frac{L^2}{3}+{2M}\right),
\quad
q=L^2\left(\frac{L^4}{27}+\frac{ML^2}{3}+\frac{Q^2}{2}\right) .
\label
{p,q_df}
\end{equation}
Substituting (\ref{p,q_df}) into (\ref{D_df}) gives
\begin{equation}
D=\frac{L^4}{27}\left[L^4\left(Q^2-M^2\right)+L^2\left(9Q^2-8M^2\right)M+ 
\frac{27}{4}\,Q^4\right].
\label
{D_sbst}
\end{equation}

If $D>0$, then there is a single real root, which, however, is negative, implying the absence of 
horizon.
If $D<0$, then there are three different real roots, one of them is negative, and two other are 
different positive roots.
If $D=0$, then there are two alternatives.
First, a single real root is realized for $p=q=0$, which, in view of (\ref{p,q_df}), is not the case for real $L, M$, $Q$.
Second, $\Delta(x)$ has a negative root, and a unique positive  root (two merged positive roots).
Let $D=0$.
Then the unique positive  root of  $\Delta(x)$ is 
\begin{equation}
x_{\ast}
=
-\left(\frac{L^2}{3}+\frac{q}{p}\right).
\label
{solution_double_positive}
\end{equation}
With the designations 
\begin{equation}
\lambda
=
\frac{27Q^2}{8M^2}\,,
\quad
\nu
=
\frac{M}{L^2}\,,
\quad
{\hat r}^2
=
\frac{x}{L^2}\,,
\label
{lambda,mu-df}
\end{equation}
Eq.~(\ref{solution_double_positive}) becomes
\begin{equation}
{\hat r}_{\ast}^2
=
\frac{\nu}{3}\left(\frac{3+{4}\lambda\nu}{1+6\nu}\right).
\label
{r_RN}
\end{equation}
This root represents a single event horizon which is peculiar to extremal black holes. 

The presumably positive solution of equation $D=0$, expressed in terms of $\lambda$ and $\nu$,  is 
\begin{equation}
\frac{1}{\nu}
=
\frac{4(\lambda-3)}{8\lambda-27}\left[\sqrt{81
-
\frac{\lambda^2(8\lambda-27)}{(\lambda-3)^2}}-9\right].
\label
{A_16}
\end{equation}
It is easy to check that the right side of Eq.~(\ref{A_16}) is positive for $0<\lambda<3$, that is,
for
\begin{equation}
Q^2< \frac{8}{9}\,M^2\,.
\label
{A_17}
\end{equation}
Thus, the only constraint on $L$, $M$, $Q$ resulting from the condition that a 5-dimensional 
anti-de Sitter Reissner--Nordstr{\o}m black hole is extremal is given by Eq.~(\ref{A_17}).

To gain an insight into the behavior of a Dirac particle in this background, we first 
rewrite Eq.~(\ref{Dirac_bulk}) in an equivalent form
\begin{equation}
\left[iG^\mu(x)\partial_\mu+\frac{i}{2}\!\left(\nabla_\mu G^\mu\right)\!(x)
+
e G^\mu(x)A_\mu(x)
-
m\right]\!\Psi(x)=0\,. 
\label
{Dirac_manifold}
\end{equation}                                           
Here,  $G^\mu(x)$ are the Dirac matrices in a curved manifold which are real linear combinations of 
the usual $\gamma$-matrices.
They are related to the metric of the curved manifold $g^{\mu\nu}$ via the anticommutation relations
$\{G^\mu(x),G^\nu(x)\}=2g^{\mu\nu}(x)$.
The term $\nabla_\mu G^\mu$ in (\ref{Dirac_manifold}) is the divergence with respect to the 
Levi-Civita connection.

In polar coordinates, the Dirac operator is given by
\[
i\left[h_t^{-1}\gamma^0\left(\frac{\partial}{\partial t}-i\frac{eQ}{r^2}\right)
+
\gamma^r\left(h_t\,\frac{\partial}{\partial r}
+\frac{{3h}_t}{r}+\frac{{h'}_t}{2}\right)\right]
\]
\begin{equation}
+ i\left[{\gamma^\psi}\left(\cot\psi+\frac{\partial}{\partial \psi}\right)
+{\gamma^\vartheta}\left(\frac12\,\cot\vartheta+\frac{\partial}{\partial \vartheta}\right)
+{\gamma^\varphi}\,\frac{\partial}{\partial \varphi}
\right]-m\,.
\label
{Dirac_operatot_out}
\end{equation}                                           
This expression clearly demonstrates that the wave function $\Psi$ can be separated into radial, 
angular and time factors \cite{Brill}, \cite{Greiner}, 
\cite{Cotaescu}, \cite{Belgiorno}, 
\cite{Wu} as might be expected from the fact that 
the background is static and spherically symmetric, $\phi=\exp(-iEt)R(r)\Theta(\psi,\vartheta,\varphi)$.
Here, $\phi$ is related to $\Psi$ via the general prescription of Ref.~\cite{Greiner},   
specialized to the metric (\ref{RN_AdS}),
$\phi
=
\left(h_t\right)^{1/2}r^{3/2}\sin\psi\left(\sin\vartheta\right)^{1/2}\Psi$.
The Dirac Hamiltonian $H$ is proportional to a linear combination of two Casimir functions.
One of them, denoted by $K$, is composed of Killing vectors associated with angular momenta, while 
the other, angular-independent, is responsible for mounting the dynamics on the mass shell.
We can therefore choose simultaneous eigenfunctions of $H$ and $K$.
The angular factor $\Theta(\psi,\vartheta,\varphi)$ is determined by the requirement $K\phi=\kappa\phi$,
where $\kappa$ are integral eigenvalues, $\kappa=\pm\left(\ell+\frac32\right), \ell=0,1,\ldots$
Only $\gamma^0$ and $\gamma^r$ remain explicitly in the radial equation after the operator $K$ is 
replaced by the number $\kappa$.
It is possible to apply a unitary transformation to the spinor space (generating the similarity 
transformation for the Dirac matrices) to represent  $\gamma^0$ and $\gamma^r$ by $2\times 2$ 
matrices, and the radial factor of $\phi$ by a two component spinor \cite{Brill}.
In this representation, the radial equation for a Dirac particle in an anti-de Sitter 
Reissner--Nordstr{\o}m background, with an electric potential energy ${eQ}/{r^2}$, becomes 
(cf. \cite{Cotaescu}, \cite{Belgiorno})
\begin{equation}
\left(h_t\,\frac{d}{dr}+\frac{\kappa}{r}\right)f
-
\left[{h_t}^{-1}\left(E-\frac{eQ}{r^2}\right)+m\right]g
=0\,,
\label
{Dirac-Hamiltonian_down}
\end{equation}
\begin{equation}
\left(h_t\,\frac{d}{dr}-\frac{\kappa}{r}\right)g
+
\left[{h_t}^{-1}\left(E-\frac{eQ}{r^2}\right)-m\right]f
=0\,.
\label
{Dirac-Hamiltonian_up}
\end{equation}

We use (\ref{Dirac-Hamiltonian_down}) for expressing $g$ in terms of $f$ and substitute the result 
in (\ref{Dirac-Hamiltonian_up}). 
We then eliminate the first derivative of  $f$ from the resulting second-order 
differential equation to obtain a Schr{\"o}dinger-like equation, much as
the corresponding result has been obtained for the set of equations (\ref{Dirac_radia_f})--(\ref{Dirac_radia}).
The calculation culminates in the effective potential
\[
U(r;E)
=
-
\frac{1}{2u}
\left[
\left(
-
\frac{6eQ}{r^4}
-
\frac{mz^2}{h_t^3}
+
\frac{mw}{h_t}\right)
-
\frac{4z}{h_t^2}\left(\frac{2eQ}{r^3}
+  
\frac{mz}{h_t}
\right)
+
\frac{8uz^2}{h_t^4}
+
\frac{2uw}{h_t^2}
\right]
\]
\begin{equation}
+
\frac{3}{4u^2}\left(
\frac{2eQ}{r^3}
+
\frac{mz}{h_t}
-
\frac{2uz}{h_t^2}
\right)^2
-
\frac{u(u-2mh_t)}{h_t^4}
+
\frac{\kappa }{rh_t^3}
\left[
\frac{h_t^2}{u}
\left(
\frac{2eQ}{r^3}
+
\frac{ u}{r}
+
\frac{mz}{h_t}
\right)
-
{z}
\right]
+
\frac{\kappa^2}{r^2h_t^2}\,,
\label
{RN_AdS_effective_potential}
\end{equation}
where $h_t$ is defined by Eq.~(\ref{f(r)}),
\begin{equation}
u(r^2)
=
E
-
\frac{eQ}{r^2}
+
mh_t\,,
\quad
w(r^2)
=
\frac{6M}{r^4}
-
\frac{10Q^2}{r^6}
-
\frac{1}{L^2}\,,
\quad
z(r^2)
=
\frac{2M}{r^3}
-
\frac{2Q^2}{r^5}
+
\frac{r}{L^2}\,.
\label
{z-df}
\end{equation}

\begin{figure}[htb]
\centerline{\includegraphics[height=\risheight,angle=270]{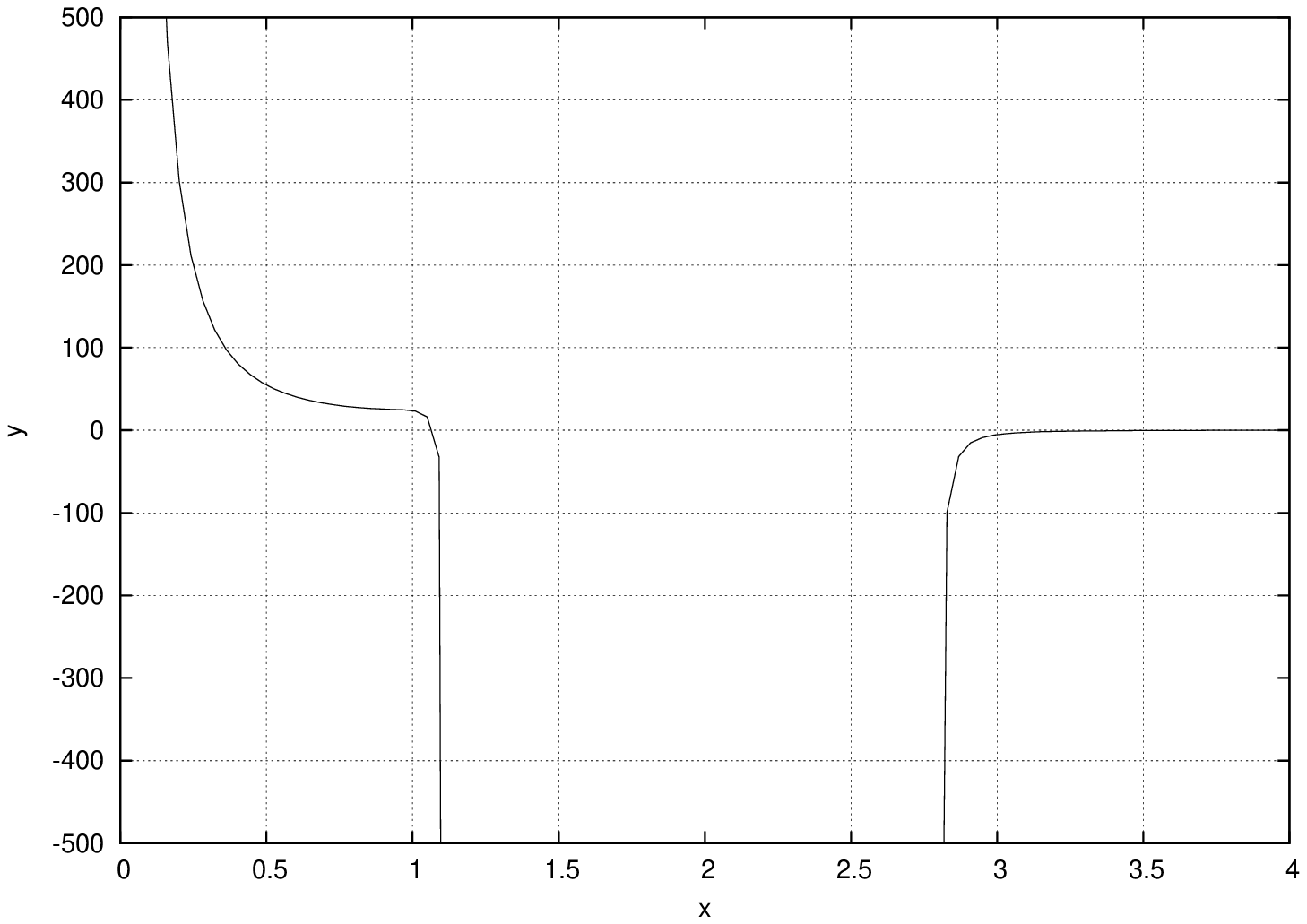}}
\caption{The effective potential (\ref{RN_AdS_effective_potential}) with the parameters $\lambda=0.210938$, 
$\nu=40$, $EL=0.1$, $mL=0.1$, $e^2/L=10^{-6}$, $\kappa=3/2$, corresponding to a non-extremal black 
hole}
\label{eff-potential-two-horiz}
\end{figure}

The effective potential $U(r;E)$ is the basic tool for probing the background.
For non-extremal black holes, $U(r;E)$ is highly singular on the outer event horizon where $h_t=0$.
The coefficient of the leading singularity is negative, so that the particle falls to the infinitely 
deep potential well at the horizon, much as a particle falls to the centre of an attractive singular potential 
\footnote{If $U(r)$ behaves near the origin as $-r^{-n},\,n\ge 2$, then one can define a 
selfadjoint Dirac Hamiltonian which exhibits a discrete spectrum extending from minus infinity to 
$m$ \cite{Case50}.
The system tends to occupy more and more advantageous states associated with 
successively lower energy levels.
As this take place, the dispersion of the wave function tends to zero as
$E\to -\infty$.},
which is most readily visualized in Fig.~\ref{eff-potential-two-horiz}. 
\psfrag{x}[c][c][0.7]{$\hat{r}$}
\psfrag{y}[c][c][0.7]{$L\cdot U(r;E)$}

\begin{figure}[htb]
\centerline{\includegraphics[height=\risheight,angle=270]{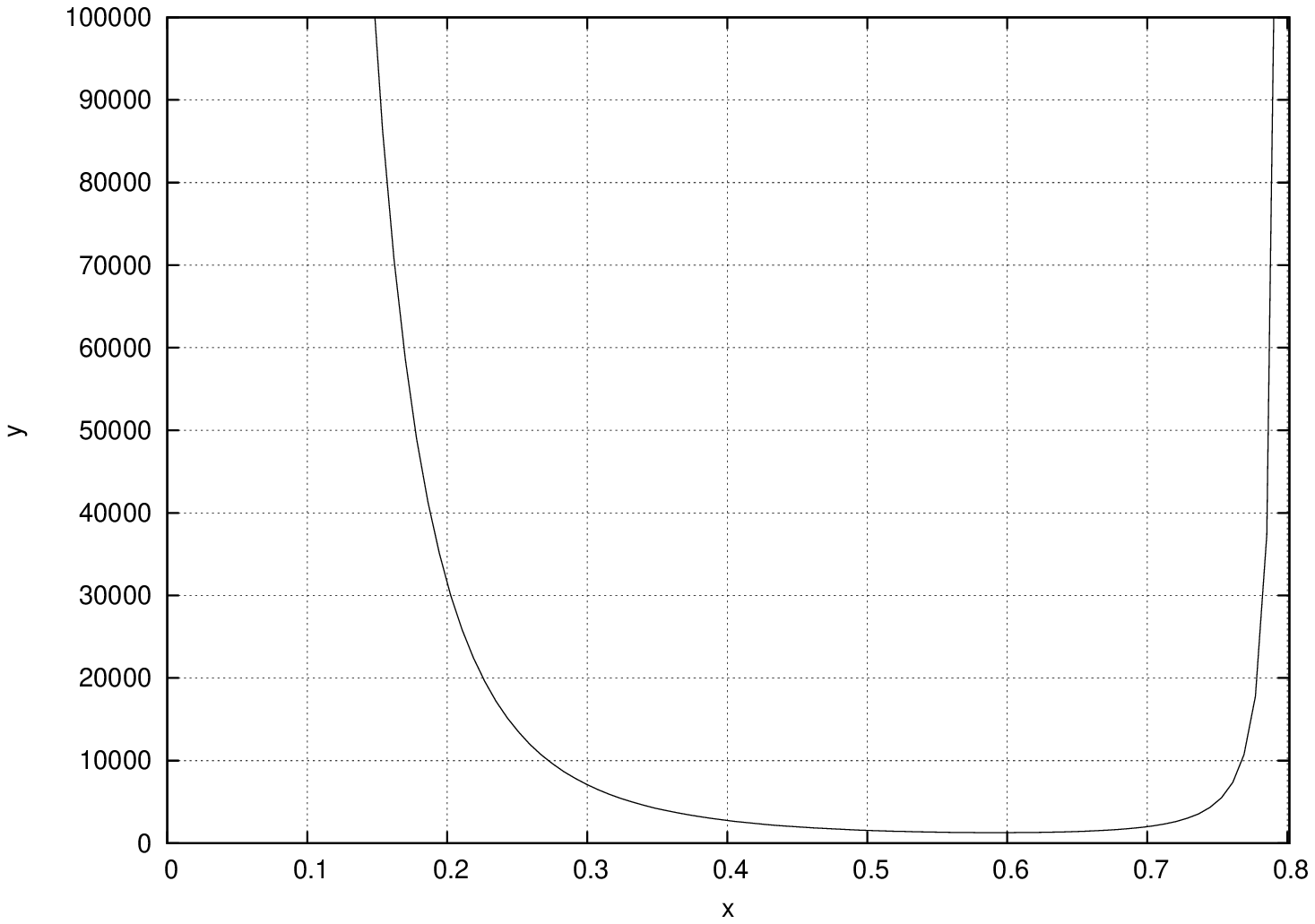}}
\caption{The effective potential (\ref{RN_AdS_effective_potential}) with the parameters $\lambda=2$, 
$\nu=1.261271$, $EL=0.015115$, $mL=0.01$, $e^2/L=10^{-4}$, $\kappa=-3/2$, corresponding to an 
extremal black hole}
\label{eff-potential-one-horiz}
\end{figure}

However, the situation can be improved if the black hole is extremal.
Indeed, one can verify that the positive double root of $\Delta(x)$ coincides with the 
positive root of $z(x)$, so that the dangerous singularities of $U(r;E)$ disappear. 
On imposing some additional condition, the coefficient of the remaining singularity becomes positive.
The pictorial rendition of the resulting $U(r;E)$ is given by Fig.~\ref{eff-potential-one-horiz}.

There is an alternative procedure which makes certain that the effective potential in the 
background of an extremal black hole can under some additional condition arrange 
itself into a smoothed infinite square well, Fig.~\ref{eff-potential-one-horiz}.
The advantage of this procedure is that it explicitly reveals this additional condition.  
Consider the behavior of a Dirac particle in the immediate vicinity of the event horizon inside an 
extremal black hole.
Anticipating that  $r_\ast$ is a turning point, that is, taking $u(r_\ast^2)=0$, we thereby fix $E$ 
to be ${eQ}/{r_\ast^2}$. 
In the limit  ${x}-{x}_\ast=-\delta\to 0$, the set of equations (\ref{Dirac-Hamiltonian_down})--(\ref{Dirac-Hamiltonian_up}) 
becomes
\begin{equation}
\frac{d}{d{\hat x}}
\pmatrix{{f} \cr
         {g}\cr}
=
\frac{\Lambda_0}{{\hat x}_\ast-{\hat x}}
\pmatrix{-\kappa & mL\sqrt{{\hat x}_\ast}-
\frac{2eQ\Lambda_0 }{L\sqrt{{\hat x}_\ast}}\cr
         mL\sqrt{{\hat x}_\ast}+
\frac{2eQ\Lambda_0 }{L\sqrt{{\hat x}_\ast}}&\kappa\cr}
\pmatrix{{f} \cr
         {g}\cr}
+O(1)\,,
\label
{truncated}
\end{equation}
where ${\hat x}_\ast$ is given by (\ref{r_RN}), and
\begin{equation}
\Lambda_0^{-1}=
{2\sqrt{3{\hat x}_\ast+1}}\,.
\label
{varpi-df}
\end{equation}
With this truncated set of equations, we reiterate mutatis mutandis the above
arguments giving rise to a Schr{\"o}dinger-like equation to conclude that
the leading term of the effective potential is proportional to 
\begin{equation}
-\frac{1}{({\hat x}-{\hat x}_{\ast})^2}
\left[\frac{1}{4\Lambda^2_0}-\kappa^2
-
{\left(mL\right)^2{\hat x}_\ast}
+
\frac{{4}\Lambda_0^2\left(eQ\right)^2}{ L^2{\hat x}_\ast}\right].
\label
{remaining_singularity}
\end{equation}
It follows that if 
\begin{equation}
{\left({m}L\right)^2}{\hat x}_\ast
-
\frac{4\Lambda_0^2(eQ)^2}{L^2{\hat x}_\ast}\ge\frac{1}{4\Lambda_0^2}-\kappa^2\,,
\label
{equilibr_condition}
\end{equation}
then $U(r;E)$ rearranges to form a smoothed infinite square well displayed in Fig.~\ref{eff-potential-one-horiz}.

Therefore, Eq.~(\ref{equilibr_condition}) represents the aforementioned additional condition.
It tells us that the Dirac particle is confined to a spherical cavity of radius $r={r}_\ast$ when 
the combined gravitational and electromagnetic influence dominate over centrifugal repulsion.
These effects are not separated but rather jumbled together in individual terms owing to the 
factors ${\hat x}_\ast$ and $\Lambda_0$ containing gravitational and electromagnetic 
contributions.
It is clear, however, that the Dirac particle is affected by gravity mostly due to the first term of 
Eq.~(\ref{equilibr_condition}), while the electromagnetic influence is attributable to the second 
term of this equation.
With this in mind, Eq.~(\ref{equilibr_condition}) is saturated by $\left(m{\cal M}\right)^2-
\left({e}{\cal Q}\right)^2 = {\cal J}^2$\,, where $m{\cal M}$ and ${e}{\cal Q}$ symbolize, 
respectively, the basic electromagnetic and gravitational contributions to Eq.~(\ref{equilibr_condition}),
and ${\cal J}^2$ is a positive number for not too great $\kappa$, or, what is the same, 
\begin{equation}
\left({m}{\cal M}-{e}{\cal Q}\right)
\left({m}{\cal M}+{e}{\cal Q}\right)
=
{\cal J}^2\,.
\label
{mathfrac_separ}
\end{equation}
Let ${e}{\cal Q}$ be positive.
This implies that the electromagnetic influence is repulsive.
We divide Eq.~(\ref{mathfrac_separ}) by the positive quantity ${m}{\cal M}+{e}{\cal Q}$ to give
\begin{equation}
{m}{\cal M}={e}{\cal Q}+{\cal C}\,,
\label
{mathfrac_eQ_positive}
\end{equation}
where ${\cal C}$ is a positive quantity, related to the centrifugal effect, which is meant for
balancing the gravitational attraction and the electromagnetic repulsion.  
In a qualitative sense, this equation has much in common with the pseudospin symmetry condition 
(\ref{PSSC}) when having regard to the fact that the impact of the attractive tensor forces of gravity is 
equivalent to that of an attractive force carried by a scalar agent.

Let then ${e}{\cal Q}$ be negative.
This implies that the electromagnetic influence is attractive.
Equation (\ref{mathfrac_separ}) is converted to
\begin{equation}
{m}{\cal M}=-{e}{\cal Q}+{\bar{\cal C}}\,,
\label
{mathfrac_eQ_negative}
\end{equation}
where ${\bar{\cal C}}$ is a positive quantity whose function is similar to that of ${\cal C}$.  
This equation resembles the spin symmetry condition (\ref{SSC}) inherent in 
free hadron states.

Note that if we take the extremal black hole background given by 
Eqs.~(\ref{RN_AdS})--(\ref{A_and_A_5}), but abandon condition (\ref{equilibr_condition}), 
then $U(r;E)$ is changed near $r=r_\ast$.
In this case, we come to the effective potential $U(r;E)$ whose analytical 
behavior evidences the following fact: if the test Dirac particle is initially 
located in a spherical cavity of radius $r_\ast$, then it will be permanently 
confined to this cavity, but if it is granted that the wave function of 
the particle is initially distributed over a region outside the cavity, then 
the particle will fall to the infinitely deep potential well at the horizon.
The form of $U(r;E)$ is very sensitive to the parameters $\lambda$, $\nu$, $E$, $m$,
$e^2$, $\kappa$, so that the run of the curve is rather irregular, and its visualization is not
illuminating.
Therefore, we omit the pictorial rendition of $U(r;E)$.

\section{Discussion and outlook}
\label
{discussion}
The Yukawa idea that the nuclear forces owe their origin to {meson exchange} mechanisms, refined by 
several innovations, such as spontaneously broken chiral symmetry, effective Lagrangians, and 
derivative expansions \cite{Weinberg1990}, forms the basis for modern nuclear physics (the present 
state of the art has been detailed in \cite{Epelbaum} and \cite{Machleidt}).
And yet the issue of understanding {nuclei in terms of quarks} is high on the agenda of the QCD 
developments.
The simplest possibility is to think of a nucleus with mass number ${\cal A}$ as a system of 
$N=3{\cal A}$ quarks placed in a bag of size $R$ \cite{Petry}.
But the stability of the bag stipulates that $N$ and $R$ must be related by $R\sim N^{1/4}$, 
contrary to (\ref{R-nucleus}), and the discord is particularly stricking for heavy nuclei. 
Furthermore, the magnetic moments of such bags significantly differ from the experimentally 
established nuclear magnetic moments \cite{Arima}, \cite{Talmi}.
An effort to account for the static properties of nuclei by eliminating gluon degrees of freedom 
was reasonably successful \cite{Maltman}, but never progressed beyond small nuclei.
The failure to utilize the bag model in nuclear physics is likely to be related to the fact that 
this model does not take proper account of {collective motion effects}. 

Another way of looking at the low energy effective theory to QCD, outlined in Sec.~2, is that a 
single quark $Q$, roaming around the nucleus, is responsible for static properties of this nucleus 
\cite{KPV}, \cite{KPV-2}.
Central to this approach is the pseudospin symmetry condition (\ref{PSSC}) applied to rising 
potentials $U_S$ and $U_V$ of the mean field generated by all degrees of freedom of the nucleus. 
The purpose of Eq.~(\ref{PSSC}) is twofold: (i) to convert the current quark mass into the 
constituent quark mass through the shift of mass, shown in (\ref{mass_shift}), and (ii) balance 
scalar attraction and vector repulsion of the mean field to confine the quark $Q$ within the nucleus.

Sound as these requirements for $U_S$ and $U_V$ may be, they are phenomenological in nature, sending 
us in search of their further substantiation.
We address a holographic mapping from the dynamical affair of a Dirac particle in a 5-dimensional 
anti-de Sitter Reissner--Nordstr{\o}m black hole to that of the quark $Q$ in a cavity representing 
a stable heavy atomic nucleus.
We find that the effective potential $U(r;E)$ developed in such gravitational manifolds never 
rearranges to form a cavity with singular boundary until the black hole becomes extremal and the 
balance condition (\ref{mathfrac_separ}) fulfils
\footnote{Note that the balance condition (\ref{equilibr_condition}) is unaffected by small quantum 
perturbations which still remain in the semiclassical, feeble quantum regime of 
evolution.
The parameters appearing in this condition have no need of fine tuning  because
we are dealing with the {inequality} which offers some range of possibilities for the perturbed 
gravitational and electromagnetic influence on a test Dirac particle to dominate over centrifugal 
repulsion.}.
Surprisingly, this balance in the bulk is an aggregate condition consistent with both pseudospin 
symmetry condition (\ref{PSSC}) and spin symmetry condition (\ref{SSC}) in the screen. 
This subject is unrelated to the main line of this paper but will hopefully be studied 
elsewhere.

This evidence in support of the assumption that a good part of nuclear physics is modelled on 
physics of extremal black holes is tempting to extend to every system of nuclear and subnuclear zoo 
for formulating a {\it necessary condition for the system to be stable}: Its {holographic 
counterpart must be extremal}.
It is interesting to apply this criterion to a clarification of the fact that truly neutral spinless 
particles (Higgs bosons, $\pi_0$, quarkonia, and glueballs \footnote{Lattice and sum rule 
calculations predict the lightest glueball to be a scalar with mass in the range of about 1 -- 1.7 
GeV \cite{Ochs}.}) are unstable.
The instability is associated with the absence of extremal objects among their counterparts, which 
are typically Schwarzschild black holes. 

It may appear that the holographic correspondence between extremal black holes and stable nuclei is 
far beyond the scope of the standard AdS/CFT which maps the AdS${}_5\times S^5$ supergravity to the 
${\cal N}=4$ super Yang--Mills theory with matter fields in the adjoint representation of the 
SU$(N_c)$ gauge group in the limit $g_{\rm YM}^2N_c\to\infty$.
The reason is that QCD is devoid of supersymmetry and conformal symmetry, and nuclear physics is an 
effective low energy theory to QCD. 

However, the same critical comment refers equally to what is presently agreed to be the established 
patterns of gauge/gravity duality. 
Take for example the holographic treatment of quark-gluon plasma in heavy ion collisions.
There are two major reasons why the dissimilarity between the symmetry 
properties of QCD and those of conformal super Yang--Mills theories may be 
erased \cite{Aref'eva}.
First, the behavior of collective excitations in the ${\cal N}=4$ super 
Yang--Mills plasma studied with the aid of the Keldysh diagram technique 
\cite{Keldysh}, in the weak coupling regime, does not significantly differ from that of Yang--Mills plasmas
devoid of supersymmetry \cite{Czajka}.
Second, since QCD is asymptotically free, it is reasonable to expect that quark-gluon plasmas at 
sufficiently high temperatures reveal scale independence, so that one may apply the equation of 
state $\epsilon=3p$ stemming from the tracelessness of the stress-energy tensor dictated by 
conformal invariance.
Lattice calculations suggest that such is the case at $T>300$ MeV \cite{Borsanyi}.

The idea that the behavior of a single quark contained in some nucleus is responsible for the static 
properties of the whole nucleus implies that collective effects (say those related to the formation 
of a spherical cavity with singular boundary which is identified with the interior of the nucleus) 
are essential for this approach.  
The only serious distinction of collective effects in a heavy nucleus from those in a quark-gluon 
plasma lump
 is that the former bear on semiclassical, feeble quantum, nonrelativistic regime of 
evolution, whereas the latter are attributable to ultrarelativistic quantum field regime.
Therefore, the erase of supersymmetry structures in imaginary ``supersymmetric nuclei'' is every 
bit as valid as that in imaginary ``supersymmetric quark-gluon plasmas''. 
As to conformal symmetry of the discussed version of gauge/gravity duality, it might be well to 
point out that the classical Yang--Mills theory in four dimension is invariant under the group of 
conformal transformations C$(1,3)$, and it is the quantum conformal anomaly that is responsible for 
violating conformal invariance of QCD (for the holographic origin of this anomaly see \cite{k2000}).
In the semiclassical, feeble quantum, contexts represented by tree Feynman diagrams, the conformal
anomaly disappears, and the missing 
conformal properties have been recovered \cite{Brodsky}.

In some respect the holographic treatment of stable heavy nuclei mimics the AdS/CFT much better 
than the holographic treatment of quark-gluon plasma copes with this task.
It is well known that quantum fluctuations disappear in the large $N_c$ limit.
To be more specific, ``for large $N_c$ the measure in function space becomes 
concentrated on a single orbit of the gauge group'', and ``the probability of 
finding any gauge invariant quantity away from its expectation value goes to 
zero as $N_c$ goes to infinity'' \cite{Coleman}.    
In other words, the 't Hooft limit $g_{\rm YM}^2N_c\to\infty$ implies that the {system in the screen 
is governed by a semiclassical, feeble quantum}, dynamics; the ``quark see'' is suppressed; and 
loop diagrams might be safely neglected.
On the other hand, the quark-gluon plasma physics is marked by creations and annihilations of 
$Q{\bar Q}$ pairs.
To realize such processes, a strongly {\it quantum} regime, in which all loop diagrams are of the
utmost significance, is called for.
But this regime is {incompatible} with the large $N_c$ limit.

In the long run, any model is assessed by its capability for unraveling still unsolved mysteries and 
puzzles, and by its predictive power.
We consider a gauge/gravity duality that offers a natural (based on the fundamental laws of gravity 
and electromagnetism) substantiation of the need for rising potentials $U_S$ and $U_V$ of the mean 
field combined with the pseudospin symmetry condition.
In addition, it furnishes insight into a striking fact that {\it all truly neutral spinless 
particles, both elementary and composite, are unstable}.  
To the best of our knowledge, the only field theoretic reason for their
instability is the nonexistence of their stable holographic counterparts.

The above general criterion for discriminating between stable and unstable systems opens a new 
avenue of attack on outstanding problems. 
For example, a surprising thing is the {\it nonexistence of stable neutral nuclei} even if a single 
neutron, regarded as the lightest nucleus of this type, is stable \footnote{Life time of a free 
neutron is about $10^3$ s which is an eternity in the standards of subnuclear realm.}.
To turn to close examination of this problem, we need solutions describing extremal rotating black 
holes in the framework of a U$(1)^2$ gauge theory parametrized by the mass $M$, two angular momenta
$J_1$ and $J_2$, and two (equal but opposite in sign) electric charges $Q_1$ and $Q_2$.
However, such manifolds are far from being completely understood \footnote{A considerable body of 
information on black holes in higher dimensions is covered in \cite{Emparan} and \cite{Horowitz}.}.
That is why we restrict our consideration to the simplest case that a Dirac particle probes the 
background of extremal Reissner--Nordstr{\o}m black holes.

It is thus seen that the holographic approach to nuclear physics can be used to 
explain certain qualitative facts about the strong interactions in the infrared.  
But we may reason in the opposite way, that is, regard the qualitative facts 
about the strong interactions as diagnostic tests showing that this approach is 
probably a good approximation to nature.

A skeptic may argue that an effort to confront black holes prone to Hawking radiation with heavy 
nuclei subjected to decay appears rather awkward because Hawking radiation is an incessant thermal 
process while the decay occurs as a single event. 
But who is to say that the disintegration of holographically duals must be similar in appearance?
The only imperative requirement is that the {mechanisms} underlying these disintegrations must 
be identical.
This is actually the case.
Both phenomena are due to quantum tunnelling~\footnote{From quantum-mechanical point of view, a 
freely evolved black hole is always in the state which is a superposition of the initial black hole 
state and the state of its remnants together with particles emitted through the Hawking evaporation 
scheme; and the same is true of a single free nucleus which is found in the state 
represented by a superposition of states of the initial nucleus and decay products.}. 
Furthermore, the {suppression of disintegration} of the discussed duals was shown to have the same
origin: both 
effective potentials,  $U(r;\varepsilon)$ and $U(r;E)$,  exhibit the singular behavior whose 
pictorial rendition is a spherical cavity with a singular boundary, and this singular wall keeps 
the Dirac particle in the cavity from escaping.

Meanwhile taking the holographic treatment of nuclear physics quite seriously, one faces a 
formidable challenge to the fundamental quantum-mechanical principle which maintains that all 
microscopic systems of a given species are {\it identical}.
For the consistency of the holography, the totality of black hole remnants must be separated into 
classes of objects with identical properties, in particular with equal masses.
On the assumption that the evaporation of black holes ends in one or more extremal black objects
(black holes, black rings, etc.), it seems incomprehensible why any history of a black hole, 
selected at random, always terminates with such classes of {identical} extremal black objects.

\section{Conclusion}
\label
{Conclusion}
The main assumption of this paper is that there is a holographic mapping between the dynamical 
affair of a single quark $Q$ in a stable heavy nucleus and that of a Dirac particle located within 
an extremal Reissner--Nordstr{\o}m black hole in 5-dimensional anti-de Sitter spacetime.
Since  semiclassical, feeble quantum, regimes of evolution is specific to both the quark $Q$ and its 
holographic dual, extremal path contributions dominate the Feynman path integrals for the partition 
functions in the bulk and in the screen.
Therefore, to contrast the dynamical affairs, it is sufficient to compare distinctive characteristics 
of a solution to the Dirac equation for a single quark $Q$ driven by the mean field of a stable 
nucleus with those of the pertinent solution to the Dirac equation in the geometry of an extremal 
black hole.
More specifically, the behavior of the effective potential $U(r;\varepsilon)$ formed in the mean 
field of the nucleus, Eq.~(\ref{U_eff-PSEUDO}), is to be confronted with the behavior of the 
effective potential $U(r;E)$ formed in the background of a black hole, Eq.~(\ref{RN_AdS_effective_potential}).  

The main result of this paper is that the form of $U(r;\varepsilon)$ bears a general resemblance to 
that of $U(r;E)$ (cf. Figures~\ref{pseudo-spin-potential} and \ref{eff-potential-one-horiz}) if it 
is granted that the mean field potentials $U_S$ and $U_V$ grow indefinitely with $r$, and obey the 
pseudospin symmetry condition (\ref{PSSC}), and, on the gravity side, the black hole is extremal, and 
the additional condition (\ref{equilibr_condition}) is met.
Equation (\ref{equilibr_condition}) signals that the attraction of gravity, electromagnetic 
influence, and centrifugal repulsion balance out, which makes the Dirac particle to be confined to a 
spherical cavity of radius $r=r_\ast$.
If the electromagnetic interaction between the black hole and the Dirac particle is repulsive, 
the additional condition (\ref{equilibr_condition}) takes the form of Eq.~(\ref{mathfrac_eQ_positive})
having much in common with the pseudospin symmetry condition (\ref{PSSC}). 

A black hole exerts on a Dirac particle by the forces derivable from the fundamental equations 
of gravitation and electromagnetism.
The pseudospin symmetry condition (\ref{PSSC}) regarded as the holographic dual of the equilibrium 
condition (\ref{mathfrac_eQ_positive}) is thus seen as not a mere phenomenological 
condition.
Therein lies the belief that the holography gives an appropriate substantiation of 
the effective theory to low energy QCD proposed in \cite{KPV}, \cite{KPV-2}.

\end{document}